\documentclass[12pt,a4paper]{article}
\pdfoutput=1
\usepackage{graphicx,epstopdf,dcolumn,bm,mathtools,siunitx,cite}
\usepackage[margin=1.0in]{geometry}
\usepackage[none]{hyphenat}
\begin{document}
\title{Characterization of density oscillations in confined and degenerate Fermi gases\thanks{NOTICE: This is an author-created version of an article accepted for publication in Modern Physics Letters B. The definitive published version is available online at doi: 10.1142/S0217984918503931.}}
\author{Coskun Firat$^{1}$, Altug Sisman$^{1,2,*}$ \& Alhun Aydin$^{1,2}$}
\maketitle
\begin{center}
\small\textit{$^{1}$Nano Energy Research Group, Energy Institute, Istanbul Technical University, 34469, Istanbul, Turkey \\
$^{2}$Department of Physics and Astronomy, Uppsala University, 75120, Uppsala, Sweden \\
$^*$Corresponding author, altug.sisman@physics.uu.se}
\end{center}
\begin{abstract} 
Friedel oscillations appear in density of Fermi gases due to Pauli exclusion principle and translational symmetry breaking nearby a defect or impurity. In confined Fermi gases, this symmetry breaking occurs also near to boundaries. Here, density oscillations of a degenerate and confined Fermi gas are considered and characterized. True nature of density oscillations are represented by analytical formulas for degenerate conditions. Analytical characterization is first done for completely degenerate case, then temperature effects are also incorporated with a finer approximation. Envelope functions defining the upper and lower bounds of these oscillations are determined. It is shown that the errors of obtained expressions are negligible as long as the system is degenerate. Numbers, amplitudes, averages and spatial coordinates of oscillations are also given by analytical expressions. The results may be helpful to efficiently predict and easily calculate the oscillations in density and density-dependent properties of confined electrons at nanoscale. 
\end{abstract}
\text{Keywords: Quantum size effects; Friedel oscillations; Nano thermodynamics}
\section{Introduction}
Global thermodynamic properties of classical and quantum gases confined in nanoscale domains have been widely considered in the literature \cite{baltes,pathriabook,molina,pathria,dai1,tse1,sisman,daixie,qbl,flopb2,uqbl,nanocav,pdx,qforce,babac,phdthesis,discnat,qosc1,qosc2}. It was shown that, global thermodynamic properties of gases confined in nanodomains are strongly affected by the sizes of the domain due to wave nature of particles. These effects are called as quantum size effects (QSE). In order to understand physical mechanisms of size dependence of global thermodynamic properties as well as to be able to use the models based on local equilibrium assumption, it is important to examine local properties, like density distribution, of confined gases. 

Local density distribution of an ideal Maxwell-Boltzmann (MB) gas confined in a rectangular domain has been shown to be not homogeneous even at thermodynamic equilibrium \cite{qbl}. Density goes to zero near to domain boundaries and a quantum boundary layer (QBL) occurs where the homogeneity in density distribution is disrupted. The thickness of this layer is on the order of thermal de Broglie wavelength of particles, and QBL vanishes when Planck's constant goes to zero. It has been seen that existence of QBL is crucial to understand and even predict QSE on thermodynamic behaviors of the particles confined in nanodomains where the particle density is not uniform. QSE terms in expressions of thermodynamic properties can directly be recovered by using QBL concept without even solving Schr\"odinger equation\cite{qbl,uqbl,nanocav}.

QSE become significant at nanoscale and bring in some unique features for nanomaterials. Semiconductor or metal nanostructures are some of the most common and convenient materials to examine QSE. Although electrons confined in non-degenerate semiconductors can be modeled by MB statistics, it is necessary to consider Fermi-Dirac (FD) statistics when dealing with confined and degenerate electron gas (\textit{e.g.} degenerate semiconductors or metals).

Unlike unbounded Fermi gases, density oscillations appear in bounded Fermi gases because of translational symmetry breaking near to boundaries in confined systems. In this study, oscillations in local density distribution of a confined and degenerate Fermi gas are examined. These density oscillations actually correspond to Friedel oscillations in metals or semiconductors due to defects or impurities. By invoking some mathematical tools such as Poisson summation formula, we first obtain an analytical expression representing the true nature of oscillations for completely degenerate case (\textit{e.g.} $T=0$K) which is called $0^{\text{th}}$ order approximation. Upper and lower envelope functions giving the ultimate bounds of the oscillations as well as amplitudes and averages of oscillations are derived. Furthermore, by making a more precise approximation, which is called $1^{\text{st}}$ order approximation, we consider the effect of temperature and derive analytical expressions also for finite temperatures. We compare exact and analytical expressions based on two different approximations and show that the errors of analytical expressions are quite low, as long as degeneracy is sufficiently high where the oscillations are considerable. Averages, amplitudes, spatial coordinates and the numbers of these oscillations are also analytically given. 

\section{Density distribution of confined and degenerate Fermi gases}
For an ideal Fermi gas in thermodynamic equilibrium, the number of particles in a differential local volume $dV$ centered at $\textbf{r}-$location for a quantum state $s$ is written as $dN_s=N\left(f_s/\sum_s{f_s}\right)|\psi_s(\textbf{r})|^2dV$, where $N$ is total number of particles, $\psi_s$ is the eigenfunction of the quantum state $s$ and $f_s=1/\left[\exp(-\Lambda+\tilde{\varepsilon}_s)+1\right]$ is FD distribution function. Here, $\Lambda=\mu/k_BT$ is degeneracy parameter, $\tilde{\varepsilon}_s=\varepsilon_s/k_BT$ is dimensionless energy eigenvalue of quantum state $s$, $\mu$ is chemical potential, $k_B$ is Boltzmann constant and $T$ is temperature. We don't take spin degree of freedom into account, since it cancels in density expression in the absence of an external magnetic field. Note that $\left(f_s/\sum_s{f_s}\right)$ denotes the probability of finding a particle in quantum state $s$ (thermodynamic probability) and $\left(|\psi_s(\textbf{r})|^2dV\right)$ gives the possibility of a particle to be in a volume $dV$ centered at $\textbf{r}-$position (quantum probability). By summing up $dN_s$ over all quantum states, local particle density reads
\begin{equation}
n(\textbf{r})=N\frac{\sum_s f_s |\psi(\textbf{r})|^2}{\sum_s f_s}=N\left\langle|\psi(\textbf{r})|^2\right\rangle_{ens}
\end{equation}
which is actually the ensemble average of quantum probability density times the number of particles.

\subsection{Density oscillations of a 1D Fermi gas}
Now let's consider a Fermi gas confined in a 1-dimensional confinement domain. The solution of Schr\"odinger equation for this kind of domain with Dirichlet boundary conditions gives dimensionless energy eigenvalues as $\tilde{\varepsilon}=(\alpha i)^2$ where $i$ is the quantum state variable running from 1 to $\infty$ and $\alpha=h/\left(\sqrt{8m k_B T}L\right)$ is confinement parameter, which denotes the strength of confinement in the domain with length $L$ and particles having mass $m$. Corresponding eigenfunctions are simply given as $\psi(x)=\sqrt{2/L}\sin(\pi xi/L)$, where $x$ denotes the position in 1D domain.

In dimensionless form, local density distribution of a Fermi gas confined in a 1D domain is represented as
\begin{equation}
\tilde{n}=\frac{n(\tilde{x})}{n_{cl}}=2\frac{\sum_{i=1}^{\infty}\frac{\sin(\pi\tilde{x}i)^2}{\exp[-\Lambda+(\alpha i)^2]+1}}{\sum_{i=1}^{\infty}\frac{1}{{\exp[-\Lambda+(\alpha i)^2]+1}}}
\end{equation}
where $n_{cl}=N/L$ is classical density and $\tilde{x}=x/L$ is dimensionless position. By using Eq. (2), we can examine the exact density distribution of an ideal Fermi gas confined in a 1D rectangular domain.

\begin{figure}[t]
\centering
\includegraphics[width=0.9\textwidth]{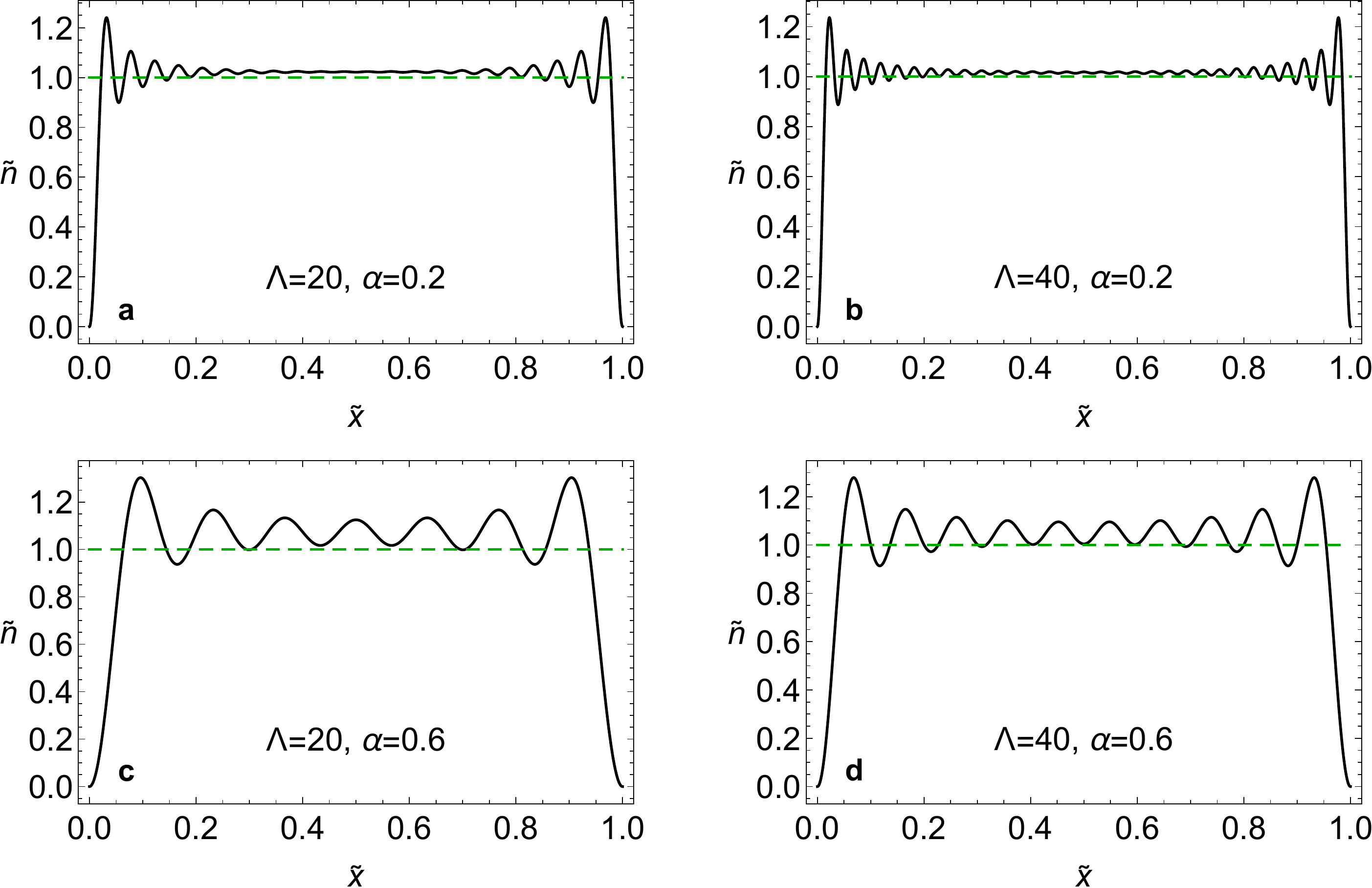}
\caption{Oscillations in dimensionless density distributions of a confined and degenerate 1D Fermi gas. Black curves represent the exact density distributions, whereas the dashed green lines show the classical density which is uniform. (\textbf{a}) $\Lambda=20, \alpha=0.2$ (relatively weaker degeneracy and confinement), (\textbf{b}) $\Lambda=40, \alpha=0.2$ (stronger degeneracy and weaker confinement), (\textbf{c}) $\Lambda=20, \alpha=0.6$ (weaker degeneracy and stronger confinement), (\textbf{d}) $\Lambda=40, \alpha=0.6$ (stronger degeneracy and confinement).}
\label{fig:pic1}
\end{figure}

Dimensionless density distributions of a confined and degenerate 1D Fermi gas are shown in Fig. 1 for four different confinement and degeneracy conditions. It is seen that increment in degeneracy leads to denser oscillations (higher wavenumber) but weakens their amplitudes. Increasing confinement, on the other hand, decrease the wavenumbers of oscillations but strengthens the oscillation amplitudes. In this sense, degeneracy and confinement have opposite effects on density oscillations.

As is seen from Fig. 1, density goes to zero near to the boundaries which shows the existence of QBL in Fermi gases. Decreasing degeneracy or increasing confinement enlarges the thickness of QBL, which means particles are affected more by the presence of boundaries in those cases. On the contrary, weak confinement and high degeneracy leads to sharper peaks that are nearest to boundaries. Although oscillations practically diminish to the middle of the domain for weak confinements, oscillations in fact become persistent even at the middle regions of the domain when confinement is relatively high.

These oscillations occurred in degenerate Fermi gases are called Friedel oscillations. In general, Friedel oscillations arise due to translational symmetry breaking nearby a defect or impurity in the system and they have an exponential decay characteristic away from symmetry breaking point \cite{mahan,frie15,nanobook,compbook,phdthesis2}. Coulomb interactions between Fermions in the system also cause Friedel oscillations \cite{frie98,frie16,frie17,frie18}. Here the same oscillations occur due to symmetry breaking caused by domain boundaries.

\subsection{Charaterization of density oscillations by $0^{\text{th}}$ order approximation}
For exact density distribution given by Eq. (2), it is not possible to obtain analytical expression without making any approximation. To obtain analytical characterization of density oscillations, we need to make approximations on FD distribution function. $0^{\text{th}}$ and $1^{\text{st}}$ order approximations to FD distribution function can be done by representing the distribution function as Heaviside step function ($\Theta$) and piecewise ramp function respectively as follows:
\begin{subequations}
\begin{align}
f=\frac{1}{\exp[-\Lambda+(\alpha i)^2]+1}&\approx f_0 =\Theta[\Lambda_F-(\alpha i)^2] \\
&\approx f_1=
\begin{cases}
	1   &, \;\; i\leq i_{min} \\
	\frac{\Lambda+1}{2}-\frac{\alpha\sqrt{\Lambda}}{2}i   &, \;\; i_{min}<i<i_{max} \\
	0   &, \;\; i\geq i_{max}
\end{cases}
\end{align}
\end{subequations}
where $\Lambda_F$ corresponds to Fermi energy, $i_{min}$ $(-)$ and $i_{max}$ $(+)$ equal to $\sqrt{\Lambda}/\alpha\mp 1/(\alpha\sqrt{\Lambda})$ respectively. Comparison of two approximations can be seen in Fig. 2.

\begin{figure}[t]
\centering
\includegraphics[width=0.55\textwidth]{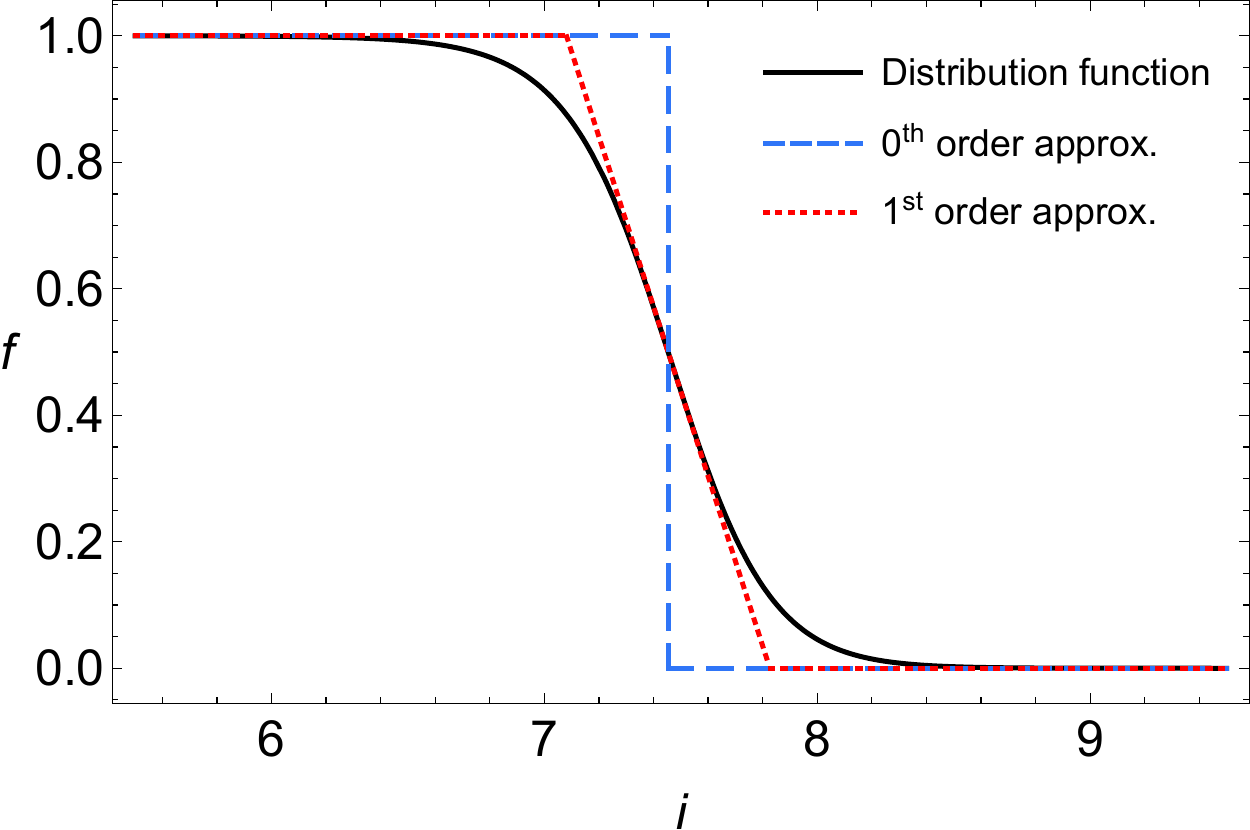}
\caption{FD distribution function is given by black curve, while dashed-blue and dotted-red curves represent $0^{\text{th}}$ and $1^{\text{st}}$ order approximations to FD distribution function respectively.}
\label{fig:pic2}
\end{figure}

We first approach to the problem for completely degenerate case, $\Lambda\rightarrow \infty$ ($T\rightarrow 0$ or $n\rightarrow\infty$), which corresponds to the $0^{\text{th}}$ approximation to FD distribution function. Under this condition ($\Lambda\rightarrow\Lambda_F$), the nominator of Eq. (2) can then be approximated using Eq. (3a) as
\begin{equation}
\begin{multlined}
\sum_{i=1}^{\infty}\frac{\sin(\pi\tilde{x}i)^2}{\exp[-\Lambda+(\alpha i)^2]+1}\xrightarrow{\Lambda\rightarrow\infty}\sum_{i=1}^{\infty}\sin(\pi\tilde{x}i)^2\Theta[\Lambda_F-(\alpha i)^2] \\
=\frac{\sqrt{\Lambda_F}}{2\alpha}\left[1-\frac{\alpha}{2\pi\sqrt{\Lambda_F}}\frac{1}{\tilde{x}}\sin\left(\frac{2\pi\sqrt{\Lambda_F}}{\alpha}\tilde{x}\right)\right]\left[1-\frac{\alpha}{2\pi\sqrt{\Lambda_F}}\frac{1}{1-\tilde{x}}\sin\left(\frac{2\pi\sqrt{\Lambda_F}}{\alpha}(1-\tilde{x})\right)\right]
\end{multlined}
\end{equation}
and the denominator can be calculated by considering the first two terms of Poisson summation formula as
\begin{equation}
\sum_{i=1}^{\infty}\frac{1}{\exp[-\Lambda+(\alpha i)^2]+1} \xrightarrow{\Lambda\rightarrow\infty}\frac{\sqrt{\Lambda_F}}{\alpha}-\frac{1}{2}
\end{equation}
Considering Eqs. (4) and (5) and defining $\alpha_F=\alpha/\sqrt{\Lambda_F}$ for brevity, fully analytical expression for Eq. (2) in completely degenerate case is obtained as
\begin{equation}
\tilde{n}_0=\frac{\left[1-\frac{\alpha_F}{2\pi}\frac{1}{\tilde{x}}\sin\left(\frac{2\pi}{\alpha_F}\tilde{x}\right)\right]\left[1-\frac{\alpha_F}{2\pi}\frac{1}{1-\tilde{x}}\sin\left(\frac{2\pi}{\alpha_F}(1-\tilde{x})\right)\right]}{1-\frac{\alpha_F}{2}}
\end{equation}
It can easily be seen from Eq. (6) that $\alpha_F$ is actually equal to the oscillation wavelength in position space. As long as $\alpha_F<<1$ (which is the condition for highly oscillatory regime), percentage errors of the analytical expression are quite low. Even for $\alpha_F=0.1$ value, the error is under $3\%$ and for $\alpha_F=0.01$ it drops to $0.15\%$. 

From Eq. (6), it is also possible to find the envelope function of the oscillations. Oscillations are direct result of the $sin$ terms in Eq. (6). Therefore, by taking the minimum and maximum values of the $sin$ function ($\mp 1$) upper ($+$) and lower ($-$) envelope functions are analytically obtained as
\begin{equation}
\tilde{n}_0^{\pm}=\frac{\left(1\pm\frac{\alpha_F}{2\pi}\frac{1}{\tilde{x}}\right)\left(1\pm\frac{\alpha_F}{2\pi}\frac{1}{1-\tilde{x}}\right)}{1-\frac{\alpha_F}{2}}\left[1\mp\frac{\alpha_F}{2\pi}\cos\left(\frac{2\pi}{\alpha_F}\right)\right]
\end{equation}
The first and second terms with brackets represent the left and right parts of the envelopes respectively. The term with square brackets represents the contribution of the counterpart when left or right part at its maximum.

Generating the envelope function for the oscillations is important as the difference of upper and lower envelopes give the ultimate bounds of the oscillation amplitude in a confined system. Thus, for completely degenerate case, bound of the oscillation amplitude depending on the position is analytically expressed as
\begin{equation}
\alpha_F<<1\;\Rightarrow\;A_0(\tilde{x})=\tilde{n}_0^{+}-\tilde{n}_0^{-}\approx\frac{\alpha_F}{\pi}\left[\frac{1}{x(1-x)}-\cos\left(\frac{2\pi}{\alpha_F}\right)\right]
\end{equation}
From the first peak to the last one, envelope functions characterize the oscillations. Before the first peak or after the last peak, envelope functions have no use. Therefore, integral average of oscillations can be found by considering the envelope functions as follows
\begin{equation}
\left\langle \tilde{n}_0\right\rangle_{osc}=\frac{\int_{\tilde{x}_1^{+}}^{1-\tilde{x}_1^{+}}{\frac{\tilde{n}_0^{+}+\tilde{n}_0^{-}}{2}d\tilde{x}}}{\int_{\tilde{x}_1^{+}}^{1-\tilde{x}_1^{+}}{d\tilde{x}}}\xrightarrow{\alpha_F\rightarrow 0}1+\frac{\alpha_F}{2}
\end{equation}
where $osc$ subscript indicates that the integral average is taken over the oscillation range. $\left\langle \tilde{n}\right\rangle_{osc}$ is actually equal to the average of envelope functions at $\tilde{x}=0.5$ (the middle point of the domain) for completely degenerate case.

For a given set of $\Lambda$ and $\alpha$, it is also possible to define the numbers of maxima and minima (numbers of peaks and dips) of oscillations which are,
\begin{subequations}
\begin{align}
& N_{peak}=\left\lfloor\frac{1}{\alpha_F}\right\rfloor \\
& N_{dip}=\left\lfloor\frac{1}{\alpha_F}-1\right\rfloor
\end{align}
\end{subequations}
where $\left\lfloor\cdots\right\rfloor$ bracket denotes the floor function. Sum of $N_{peak}$ and $N_{dip}$ give number of extremum points of oscillations. Then, positions of maxima and minima of oscillations can also be found as
\begin{equation}
\tilde{x}_j^{\pm}=\frac{4j\mp 1}{4}\alpha_F, \mbox{   with   } {j}=1,2,3,\ldots,j_{max}
\end{equation}
where $j_{max}=N_{peak}/2$ for maxima ($+$) and $j_{max}=N_{dip}/2$ for minima ($-$). Although Eq. (11) is valid only for left part ($0\leq\tilde{x}\leq 0.5$), the right part can easily be found by $1-\tilde{x}_j^{\pm}$, which is valid between the interval $0.5\leq\tilde{x}\leq 1$. Density values corresponding to extrema of oscillations can easily be found by replacing $\tilde{x}$ in Eq. (6) with Eq. (11). Relative percentage error of Eq. (11) with respect to Eq. (6) is under $5\%$ for the first peak and decrease sharply below $1\%$ right after.

\subsection{Charaterization of density oscillations by $1^{\text{st}}$ order approximation}

Even though $0^{\text{th}}$ order analysis gives some analytical expressions with reasonable errors, these results are obtained just for zero temperature. In other words, $\alpha_F$ does not depend on temperature and the results of $0^{\text{th}}$ order approximation does not represent temperature dependence. In order to incorporate temperature into results and to get higher precision, we can approximate FD in a finer way using the piecewise ramp function defined in Eq. (3b).

Using the similar methodology that is followed in the derivation of $0^{\text{th}}$ order expressions, we can use Eq. (3b) this time to calculate the density distribution. After mathematical operations the expression is simplified as
\begin{equation}
\begin{multlined}
\tilde{n}_1=\frac{\left[1-\frac{\alpha^2}{4\pi^2}\frac{1}{\tilde{x}^2}\sin\left(2\pi\frac{\sqrt{\Lambda}}{\alpha}\tilde{x}\right)\sin\left(2\pi\frac{1}{\alpha\sqrt{\Lambda}}\tilde{x}\right)\right]}{1-\frac{\alpha}{2\sqrt{\Lambda}}} \\
\times\left[1-\frac{\alpha^2}{4\pi^2}\frac{1}{(1-\tilde{x})^2}\sin\left(2\pi\frac{\sqrt{\Lambda}}{\alpha}(1-\tilde{x})\right)\sin\left(2\pi\frac{1}{\alpha\sqrt{\Lambda}}(1-\tilde{x})\right)\right]
\end{multlined}
\end{equation}
Considering the minimum and maximum values of trigonometric products in Eq. (12), envelope functions for the $1^{\text{st}}$ order approximation are found as
\begin{equation}
\begin{multlined}
\tilde{n}_1^{\pm}=\frac{\left[1\pm\frac{\alpha^2}{4\pi^2}\frac{1}{\tilde{x}^2}\sin\left(2\pi\frac{1}{\alpha\sqrt{\Lambda}}\tilde{x}\right)\right]\left[1\pm\frac{\alpha^2}{4\pi^2}\frac{1}{(1-\tilde{x})^2}\sin\left(2\pi\frac{1}{\alpha\sqrt{\Lambda}}(1-\tilde{x})\right)\right]}{1-\frac{\alpha}{2\sqrt{\Lambda}}} \\
\times\left[1\mp\frac{\alpha^2}{4\pi^2}\cos\left(2\pi\frac{\sqrt{\Lambda}}{\alpha}\right)\sin\left(2\pi\frac{1}{\alpha\sqrt{\Lambda}}\right)\right]
\end{multlined}
\end{equation}
From the difference of upper and lower envelopes given by Eq. (13), oscillation amplitude in the $1^{\text{st}}$ order approximation are found as
\begin{equation}
\alpha<<1\;\Rightarrow\;A_1(\tilde{x})=\tilde{n}_1^{+}-\tilde{n}_1^{-}\approx\frac{\alpha^2}{2\pi^2}\left[\frac{\sin\left(2\pi\frac{1}{\alpha\sqrt{\Lambda}}\tilde{x}\right)}{\tilde{x}^2}+\frac{\sin\left(2\pi\frac{1}{\alpha\sqrt{\Lambda}}(1-\tilde{x})\right)}{(1-\tilde{x})^2}\right]
\end{equation}
Oscillation average is given as
\begin{equation}
\left\langle \tilde{n}_1\right\rangle_{osc}=\frac{\int_{\tilde{x}_1^{+}}^{1-\tilde{x}_1^{+}}{\frac{\tilde{n}_1^{+}+\tilde{n}_1^{-}}{2}d\tilde{x}}}{\int_{\tilde{x}_1^{+}}^{1-\tilde{x}_1^{+}}{d\tilde{x}}}\xrightarrow{\alpha\rightarrow 0}1+\frac{\alpha}{2\sqrt{\Lambda}}+\left(\frac{\alpha}{2\sqrt{\Lambda}}\right)^2
\end{equation}
Analytical density distribution, envelope functions, amplitude and average of oscillations give better results than their $0^{\text{th}}$ order counterparts respectively as expected. Note that number of oscillations stays the same in the $1^{\text{st}}$ order approximation also, so Eq. (10) does not change. On the other hand, it is not possible to give the positions of oscillation peaks and dips analytically in the $1^{\text{st}}$ order approximation. They can be numerically obtained from the solution of $\tilde{x}$ from $\cot(2\pi\tilde{x}/\alpha\sqrt{\Lambda})+\Lambda\cot(2\pi\tilde{x}\sqrt{\Lambda}/\alpha)=\frac{\alpha\sqrt{\Lambda}}{\pi\tilde{x}}$.

Comparison of exact and analytical expressions as well as the accuracy of envelope functions are shown in Fig. 3 for various degeneracy and confinement values. Black, dashed-blue and dashed-red curves represent the results of exact (Eq. 2), $0^{\text{th}}$ order analytical (Eq. 6) and $1^{\text{st}}$ order analytical (Eq. 12) expressions respectively. Envelope functions are given by dashed-gray curves. For each figure, the upper subfigures are the results of $0^{\text{th}}$ order approximation, while the lower ones are the $1^{\text{st}}$ order ones. It is seen that analytical expressions accurately represents the true nature of density oscillations. Accuracies of envelope functions given by Eq. (7) ($0^{\text{th}}$ order) and Eq. (13) ($1^{\text{st}}$ order) in describing the upper and lower bounds of the oscillations are quite well for all cases. 

\begin{figure}[t]
\centering
\includegraphics[width=0.95\textwidth]{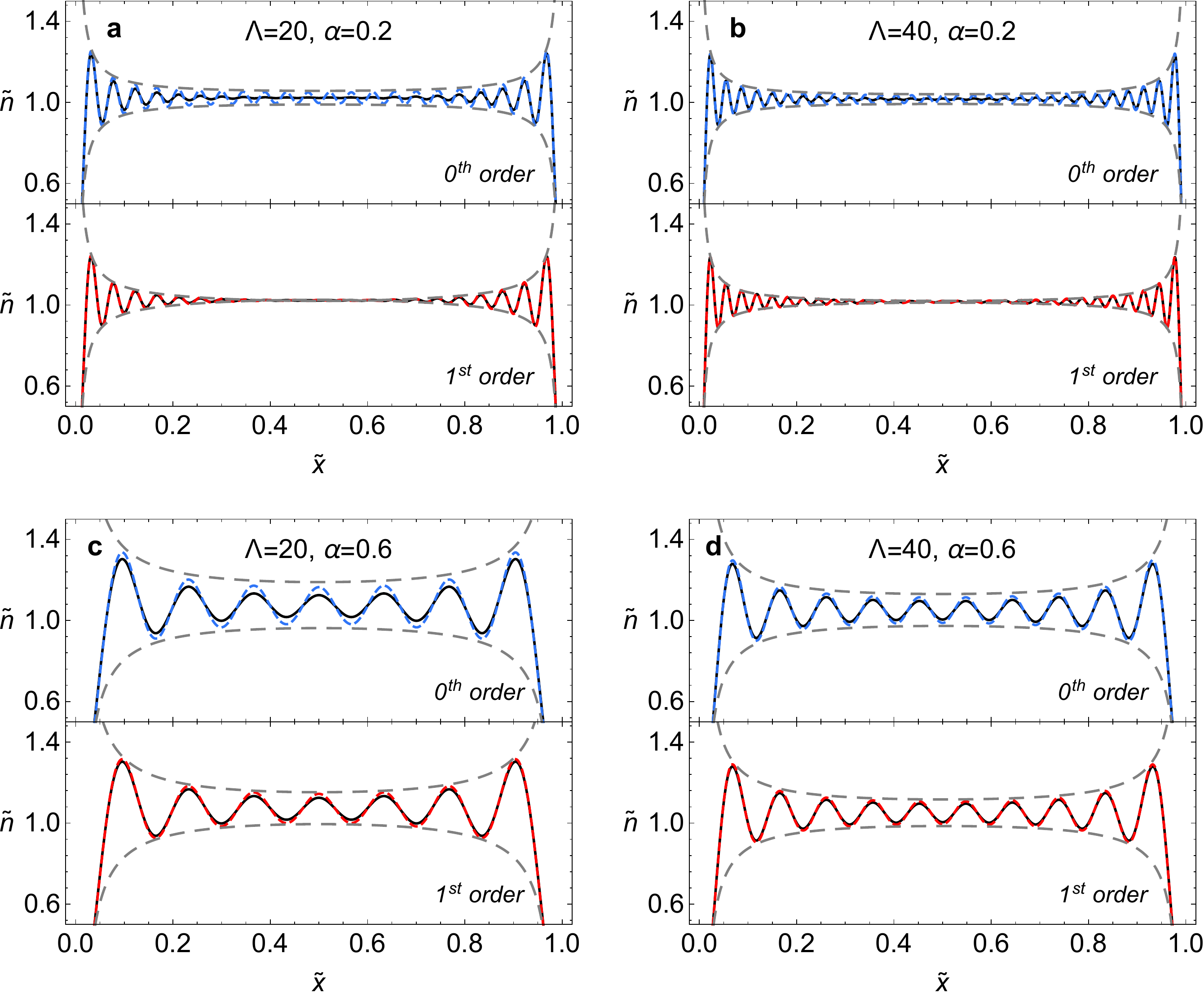}
\caption{Comparison of analytical characterization of Friedel oscillations in dimensionless density distributions of a confined and degenerate 1D Fermi gas. Black, dashed-blue and dashed-red curves represent the results of exact, $0^{\text{th}}$ order and $1^{\text{st}}$ order analytical expressions respectively. Upper and lower dashed-gray envelope functions passing from the extrema of the analytical function define the bounds of the oscillations analytically. (\textbf{a}) $\Lambda=20, \alpha=0.2$, (\textbf{b}) $\Lambda=40, \alpha=0.2$, (\textbf{c}) $\Lambda=20, \alpha=0.6$, (\textbf{d}) $\Lambda=40, \alpha=0.6$.}
\label{fig:pic3}
\end{figure}

Although both approximations represent the oscillations quite good, the results of $1^{\text{st}}$ order approximation are naturally better. Comparison of both approximations can more directly be seen in Fig. 4, where relative differences of both approaches are plotted for four different degeneracy-confinement values used in Fig. 3. $1^{\text{st}}$ order approximation always has lower errors than the $0^{\text{th}}$ one and it gives particularly good results when confinement and/or degeneracy is weak. For strongly degenerate cases, both approximations start to approach each other. Since $1^{\text{st}}$ order approximation accurately takes the slope of FD distribution around Fermi level, temperature appears in equations, particularly by the $\alpha\sqrt{\Lambda}$ term.

\begin{figure}[t]
\centering
\includegraphics[width=0.65\textwidth]{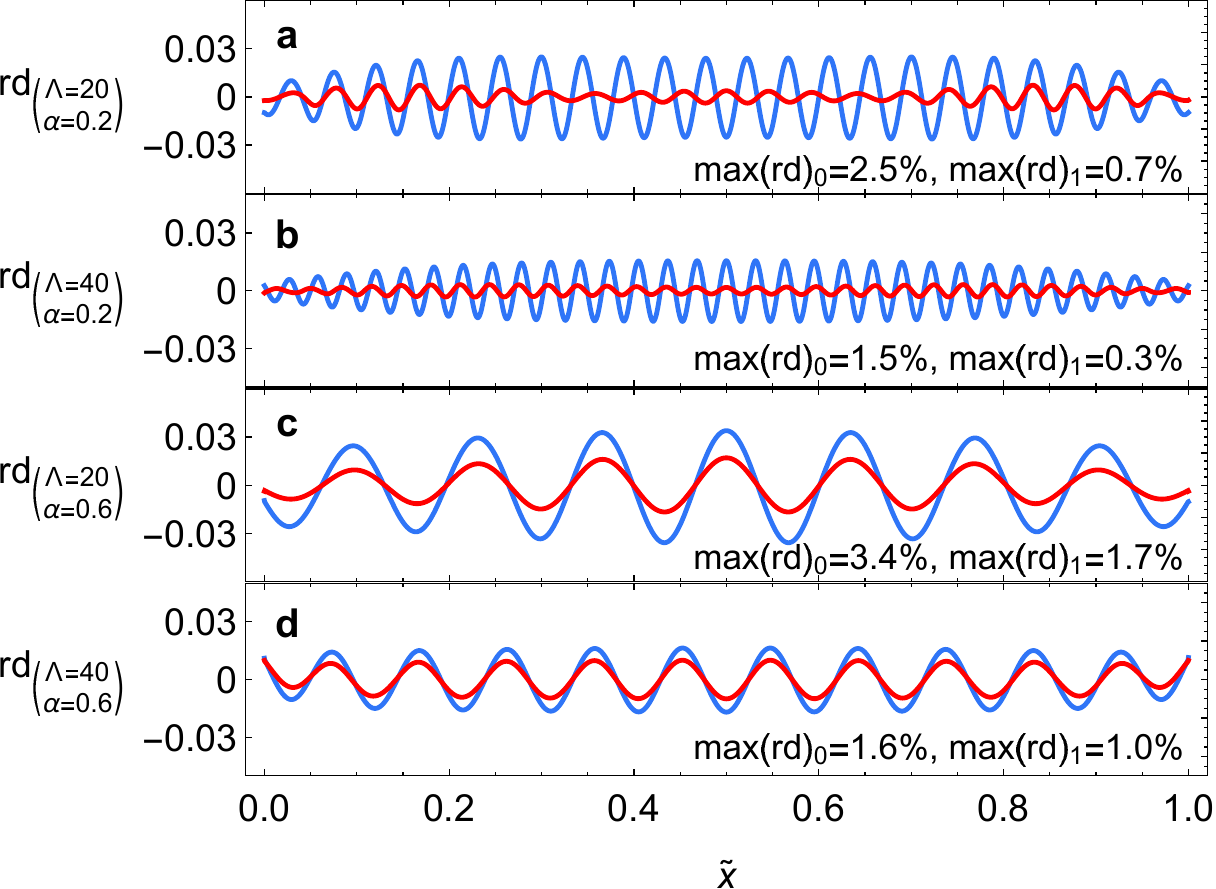}
\caption{Relative differences of $0^{\text{th}}$ order (blue curves) and $1^{\text{st}}$ order (red curves) approximations. For various degeneracy and confinement conditions, (\textbf{a}) $\Lambda=20, \alpha=0.2$, (\textbf{b}) $\Lambda=40, \alpha=0.2$, (\textbf{c}) $\Lambda=20, \alpha=0.6$, (\textbf{d}) $\Lambda=40, \alpha=0.6$, maximum relative error values of $0^{\text{th}}$ and $1^{\text{st}}$ order density expressions are given inside each sub-figure.}
\label{fig:pic4}
\end{figure}

\section{Conclusion}
In this study, we considered a degenerate Fermi gas confined in a 1D domain and examined the characteristics of density oscillations by making $0^{\text{th}}$ order and $1^{\text{st}}$ order approximations to FD distribution function. By using mathematical tools like Poisson summation formula, we obtained accurate analytical expressions for both the density profile and its characteristic parameters such as oscillation numbers, amplitudes, averages and their spatial coordinates. Although $0^{\text{th}}$ order approximation does not take the effects of temperature into account, with the $1^{\text{st}}$ order approximation, temperature is also incorporated into the results.

It should be noted that although the characterization of density oscillations is done by considering 1D Fermi gas, it is trivial to consider 2D and 3D cases. Due to the orthogonality of eigenstates, the analytical expressions obtained in this article can be directly extended into higher dimensions in a rectangular confinement domain, which is actually a common geometry in semiconductor structures. On the other hand, the methods that are used in this article take advantage from the characteristic nature of Fermi-Dirac distribution function, so it can only be used to characterize the degenerate and confined Fermi gases.

With the help of the expressions given here, it may be possible to efficiently and easily predict and calculate the density oscillations as well as the oscillations in density-dependent properties of confined Fermions.
\section*{Acknowledgements}
The author A.A. gratefully acknowledges support from AIM Energy Technologies Corp. Authors are indebted to the anonymous reviewer for providing insightful comments which greatly contributed to the further improvement of the article.
\bibliography{friref}

\begin{thebibliography}{10}

\bibitem{baltes}
H.~P. Baltes and E.~R. Hilf.
\newblock {\em Spectra of finite systems}.
\newblock Bibliographisches Institut, 1976.

\bibitem{pathriabook}
R.~K. Pathria and P.~D. Beale.
\newblock {\em Statistical Mechanics, 3rd Ed.}
\newblock Pergamon Press, 2011.

\bibitem{molina}
M.~I. Molina.
\newblock {\em Am. J. Phys.}, 64:503--505, 1996.

\bibitem{pathria}
R.~K. Pathria.
\newblock {\em Am. J. Phys.}, 66:1080--1085, 1998.

\bibitem{dai1}
W.~S. Dai and M.~Xie.
\newblock {\em Phys. Lett. A}, 311:340--346, 2003.

\bibitem{tse1}
A.~Sisman and I.~Muller.
\newblock {\em Phys. Lett. A}, 320:360--366, 2004.

\bibitem{sisman}
A.~Sisman.
\newblock {\em J. Phys. A- Math. Gen.}, 37:11353--11361, 2004.

\bibitem{daixie}
W.~S. Dai and M.~Xie.
\newblock {\em Phys. Rev. E.}, 70:016103, 2004.

\bibitem{qbl}
A.~Sisman, Z.F. Ozturk, and C.~Firat.
\newblock {\em Phys. Lett. A}, 362:16--20, 2007.

\bibitem{flopb2}
T.~Fulop and I.~Tsutsui.
\newblock {\em J. Phys. A: Math. Theor.}, 42:475301, 2009.

\bibitem{uqbl}
C.~Firat and A.~Sisman.
\newblock {\em Phys. Scr.}, 79:065002, 2009.

\bibitem{nanocav}
C.~Firat, A.~Sisman, and Z.F. Ozturk.
\newblock {\em Energy}, 35:814--819, 2010.

\bibitem{pdx}
H.~Pang, W.S. Dai, and M.~Xie.
\newblock {\em J. Phys. A: Math. Theor.}, 44:365001, 2011.

\bibitem{qforce}
C.~Firat and A.~Sisman.
\newblock {\em Phys. Scr.}, 87:045008, 2013.

\bibitem{babac}
A.~Sisman and G.~Babac.
\newblock {\em Continuum Mech. Thermodyn.}, 24:339--346, 2012.

\bibitem{phdthesis}
C.~Firat.
\newblock {\em Quantum size effects on the thermodynamic behavior of gases}.
\newblock PhD thesis, Istanbul Technical University, Energy Institute, 12 2007.
\newblock PhD Thesis (In Turkish).

\bibitem{discnat}
A.~Aydin and A.~Sisman.
\newblock {\em Phys. Lett. A}, 378:2001--2007, 2014.

\bibitem{qosc1}
A.~Aydin and A.~Sisman.
\newblock {\em Phys. Lett. A}, 382:1807--1812, 2018.

\bibitem{qosc2}
A.~Aydin and A.~Sisman.
\newblock {\em Phys. Lett. A}, 382:1813--1817, 2018.

\bibitem{mahan}
G.~D. Mahan.
\newblock {\em Int. J. Mod. Phys. B}, 9:1327--1341, 1995.

\bibitem{frie15}
C.~Bena.
\newblock {\em C. R. Physique}, 17:302--321, 2016.

\bibitem{nanobook}
H.~E. Schaefer.
\newblock {\em The Science of the Small in Physics}.
\newblock Springer-Verlag, 2010.

\bibitem{compbook}
D.~Neilson and M.~P. Das.
\newblock {\em Computational Approaches to Novel Condensed Matter Systems}.
\newblock Springer Science+Business, 1995.

\bibitem{phdthesis2}
V.~M. Stojanovic.
\newblock {\em Novel ordered states of matter in ultra-cold atomic gases}.
\newblock PhD thesis, Carnegie Mellon University, 2008.
\newblock PhD Thesis.

\bibitem{frie98}
Y.~Zhang Q.~Yuan, H.~Chen and Y.~Chen.
\newblock {\em Phys. Rev. B}, 58:1084--1087, 1998.

\bibitem{frie16}
E.~G.~Dalla Torre, D.~Benjamin, Y.~He, D.~Dentelski, and E.~Demler.
\newblock {\em Phys. Rev. B}, 93:205117, 2016.

\bibitem{frie17}
K.~Riechers, K.~Hueck, N.~Luick, T.~Lompe, and H.~Moritz.
\newblock {\em Eur. Phys. J. D}, 71:232, 2017.

\bibitem{frie18}
J.~M. Zhang and Y.~Lui.
\newblock {\em Phys. Rev. B}, 97:075151, 2018.

\end{thebibliography}
\bibliographystyle{unsrt}
\end{document}